\documentclass[reprint,12pt]{JASA}
\usepackage[utf8]{inputenc}
\usepackage[T1]{fontenc}
\usepackage{graphicx}
\usepackage{grffile}
\usepackage{wrapfig}
\usepackage{rotating}
\usepackage[normalem]{ulem}
\usepackage{amsmath}
\usepackage{textcomp}
\usepackage{amssymb}
\usepackage{capt-of}
\usepackage{hyperref}
\usepackage{graphicx}% Include figure files
\usepackage{cleveref}
\usepackage{siunitx}
\usepackage{float}
\usepackage{subfigure}
\usepackage{adjustbox}
\date{\today}
\title{}
\hypersetup{
 pdfauthor={thezuf},
 pdftitle={},
 pdfkeywords={},
 pdfsubject={},
 pdfcreator={Emacs 26.3 (Org mode 9.1.9)}, 
 pdflang={English}}
\begin{document}

\title{Head-related transfer function measurements in a compartment fire}

\author{Mustafa Z. Abbasi} \email{mustafa\_abbasi@utexas.edu}

\thanks{Corresponding author.}
\author{Preston S. Wilson}
\author{Ofodike A. Ezekoye}
\affiliation{Walker Department of Mechanical Engineering, The University of Texas at Austin, Austin, Texas }

\date{\today}

\begin{abstract}
The Personal Alert Safety System (PASS) is an alarm signal device carried by firefighters to help rescuers locate and extricate downed firefighters. A fire creates temperature gradients and inhomogeneous time-varying temperature, density, and flow fields that modify the acoustic properties of a room. To understand the effect of the fire on an alarm signal, experimental measurements of head-related transfer functions (HRTF) in a room with fire are presented in time and frequency domains. The results show that low frequency (<1000~Hz) modes in the HRTF increase in frequency and higher frequency modal structure weakens and becomes unstable in time. In the time domain, the time difference of arrival between the ears changes and becomes unstable over time. Both these effects could impact alarm signal detection and localization. Received level of narrowband tones is presented that shows the fire makes the received level of a source vary by >10~dB. All these effects could impact the detection and localization of the PASS alarm, and life safety consequences.
\end{abstract}

\maketitle

\section{Introduction}
\label{sec:orgea67b18}

The Personal Alert Safety System (PASS) is a device carried by firefighters, typically mounted on their Self Contained Breathing Apparatus (SCBA). The device detects motion and emits an alarm sound if the firefighter has been stationary for too long. Firefighters rely on the PASS alarm to alert others if they need to be rescued. Rescuers can follow the alarm sound to the source and extricate the downed firefighter. The PASS device and the PASS alarm signal are governed by the NFPA 1982 Standard on Personal Alert Safety Systems (PASS). The most recent standards have required an alarm signal consisting of a series of warbled tones \cite{NFPA1982-2007},  a series of frequency sweeps between 2 kHz and 4 kHz \cite{NFPA1982-2013} and a combination of tonal warbles and frequency sweeps \cite{NFPA1982-2018}.

The PASS alarm has been in service since 1983 and has been successful as a life safety device. However, anecdotal evidence (including personal communication) from the fire service and a survey of National Institute of Occupation Health and Safety (NIOSH) fatality reports suggests that the PASS does not always work as expected \cite{ford_NIOSH_PASS}.  After a firefighter fatality, NIOSH conducts a post-incident report to clarify the factors contributing to a line-of-duty death (LODD).  A survey of these reports from 1997--2011 was conducted and found that in many cases the PASS alarm was either not heard, or not localized properly, leading to adverse outcomes. This work aims to understand the physics of sound propagation on the fireground, with the final goal of improving the PASS alarm signal and reduce the failure rate.

\subsection{Head related transfer function}
\label{sec:org0a66130}
\label{sec:intro_hrtf_desc}

The head-related transfer function (HRTF) is the frequency-dependent ratio of the sound present at the apex of the ear canal to the sound at the source. Assuming linear acoustics, and a time-domain description, the received signal at the left ear \(x_l(t)\) due to a source signal \(x(t)\) can be found using the convolution
\begin{equation}
x_l(t) = \int h_l(t) x(t- \tau) d\tau,
\end{equation}
where \(h_l(t)\) is the response at the left ear due to a delta function at the source. The Fourier transform of \(h_l(t)\) yields \(H_l(f)\), the HRTF in the frequency domain. The same formulation is applied to the right ear. 

Sound source localization by humans is a complex phenomenon. Two dominant mechanisms are the interaural level difference (ILD) and the interaural time delay (ITD) \cite{alma991057934232906011}. ILD is the difference in receive level of sound source between the two ears (i.e. the ratio of HRTF to the right and left ears), mainly due to interference by the solid head. \citet{doi:10.1121/1.1500759} shows that less than 1 dB of level difference can be differentiated and used for localization. ILD is used as a localization cue above 1500 Hz \cite{WightmanFL1992Tdro}.  ITD is the time difference of arrival of a sound at a pair of ears from a single source. If the sound arrives at the right before the left, that is a sign that the source is to the right of the listener. For continuous or long-duration sounds, ITD is a low-frequency cue (< 1500 Hz). \citet{suits2013development} showed measurements of the effect of firefighter PPE on HRTFs using the KEMAR acoustic manikin.

This paper addresses the impact of the fire on the sound heard by a firefighter in the room. Head-related transfer functions (HRTF) are measured over time as the fire evolves and provide insight into how the fire would change the sound heard by the firefighter.  The head-related transfer function (HRTF) impacts the localization and detection of sound. Much of the previous literature focuses on measuring the HRTF, and the localization and source discrimination it allows. This work focuses on understanding how the dynamic acoustic environment created by the fire affects the HRTF. HRTFs were measured before ignition, while the fire was burning, and after extinction.

\section{Experimental Design}
\label{sec:orgde41005}
\label{sec:hrtf_exp_schems}

\begin{figure*}
\centering
\includegraphics[keepaspectratio ,,width=1.0\linewidth,height=1.3\textheight]{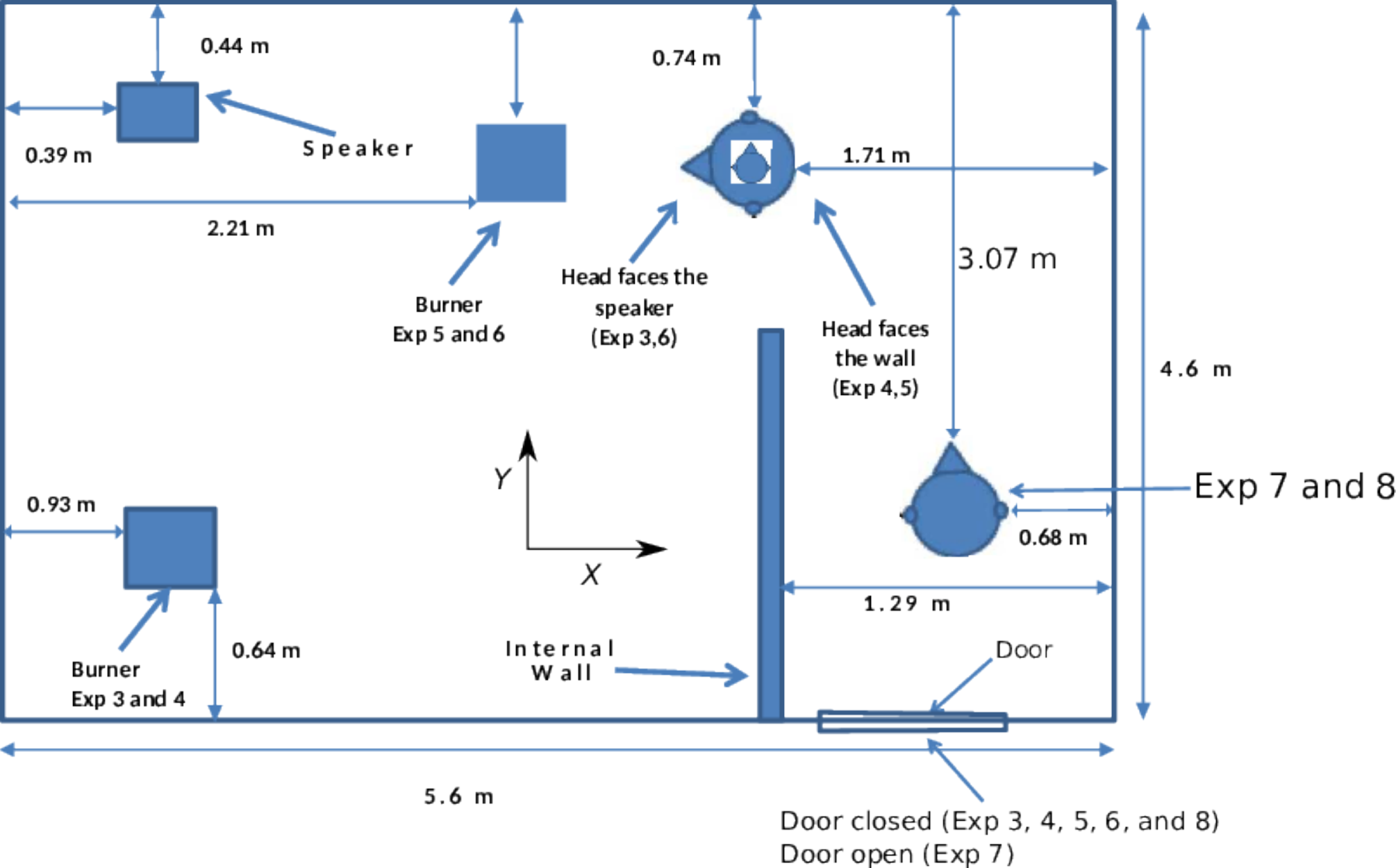}
\caption{\label{fig:exp3_schem}
(Color Online)  Top view of the burn compartment and equipment for the experiments discussed in this paper (experiments 3 through 8). The two locations of the burner, speaker, and glass manikin head are marked and labeled by experiment number. The door into the compartment is marked and was open in experiments 3 through 7, and closed in experiment 8.}
\end{figure*}

Compartment fire testing was conducted at the University of Texas burn facility. The total floor area is 5.6 m (\(X\)) x 4.6 m (\(Y\)) and the height of the ceiling is 2.1 m (\(Z\)). The room was divided into the main compartment and a hallway with a door. A series of eight experiments were designed, of which the results of six are discussed in this article. Experiment 1 is discussed in \cite{Abbasi2020_Change} and Experiment 2 is not presented because of equipment failure during that experiment. Experiments 3 through 8 use a manikin head to measure the HRTF over time as the fire developed in the room. \Cref{fig:exp3_schem} shows the positions of the speaker, burner, and manikin head in experiments 3--8. The following scenarios were modeled:
\begin{itemize}
\item The downed firefighter is in a corner of the room, opposite where the fire is burning. A rescuer is crawling on the floor while searching for the PASS beacon. This scenario is addressed in experiments 3 and 4.
\item The downed firefighter is in a corner of the room and the fire is between the firefighter and the rescuer. This scenario is addressed in experiments 5 and 6.
\item The downed firefighter is in a corner of the room, opposite where the fire is burning. A rescuer enters the open door and stays in the hallway, crawling to find the PASS beacon.  This scenario is addressed in experiment 7.
\item The rescuer enters the room like in experiment 7, but the door closes behind them. This scenario is addressed in experiment 8.
\end{itemize}
These experiments were designed to gain an understanding of the physical acoustics of the compartment fire, while still being grounded in the reality of the PASS problem.  While the above scenarios involve a rescuer crawling on the floor, the receivers were stationary during the experiments.

A 150 kW fire is used for all experimental work and is a small fire compared to those seen on a real fireground. However, larger fires also create an uninhabitable environment that rescuers would not enter. In that respect, this fire likely creates an environment representative of compartment fires where rescuers can reasonably enter and extricate a downed firefighter.

\section{Signal Processing}
\label{sec:org2be99f4}

\label{sec:impulse_response_signal_processing}

The acoustic transfer function of a source and receiver pair in a compartment fire was measured using deconvolution of a source and received signal. The source signal was a 0.2-s-long chirp, logarithmically modulated from \SI{100}{\Hz} to \SI{5000}{\Hz}. The signal was generated by a MATLAB\textsuperscript{\textregistered} program and converted to an analog signal using a NI-9263 analog output module at a \SI{100}{\kHz} sample rate. 

The instantaneous frequency \(f_i\) of the chirp was 
\begin{equation} 
\label{eq:log_frequency_chirip}
  f_i(t) = f_0 \beta^{t}, \beta = \left( \frac{f_1}{f_0} \right)^{ \frac{1}{t_1}}, 
\end{equation} 
where \(t\) is time, \(f_0\) is the start frequency, \(f_1\) is the end frequency and \(t_1\) is duration of the sweep \cite{MATLAB_2014a}. The frequency response of the system was found from the discrete Fourier transforms of the microphone signals and the input chirp as described in \cite{schafer1989discrete,muller2001transfer}. \Cref{eq:processing,eq:processing2}  describe this method: 
\begin{equation} 
\label{eq:processing}
  H(f) = \frac{\textrm{DFT}[s_{r}(t)]}{\textrm{DFT}[s_{t}(t)]}, 
\end{equation} 
and 
\begin{equation} 
\label{eq:processing2}
  h(t) = \textrm{Re}\{\textrm{IDFT}[ H(f)]\}, 
\end{equation} 
where \(s_{t}(t)\) is the transmit signal and \(s_{r}(t)\) is the received signal.  The discrete Fourier transform of \(x(t)\) is DFT[\(x(t)\)], the inverse discrete Fourier transform of \(x(f)\) is IDFT[\(x(f)\)], the real part of \(x(t)\) is Re[\(x(t)\)], \(H(f)\) is the complex frequency response of the system and \(h(t)\) is the impulse response of the system. The signal processing described in this section was used for all the experiments presented in this chapter.

The frequency and impulse response were measured repeatedly in sequence as a function of time \(T\) since ignition and will henceforth be referred to as \(H(f,T)\) and \(h(t,T)\) respectively.

The interaural level difference (ILD) is computed by \(10\log_{10}|H_r /{H_l}|^2\) where \(H_r\) is the frequency response of the right ear microphone and \(H_l\) is the frequency response of the left ear microphone.  The Interaural time delay is computed using the cross-correlation of impulse responses at the right and left ear.

\section{Head-Related Transfer Function Measurement}
\label{sec:org4fcb637}

\subsection{Experimental setup and apparatus}
\label{sec:orgd5d50f3}
\label{sec:experiment_2_overview}

\begin{figure}[]
\centering
\includegraphics[width=0.9\linewidth]{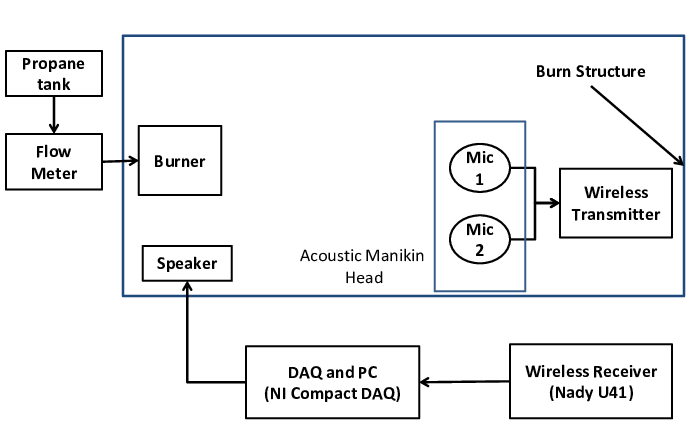}
\caption{\label{fig:schematic1_hrtf_measurement}
(Color Online) Block diagram for the time/frequency impulse response measurement in experiments 3, 4, 5, 6, 7, 8. A sand burner is used to contain the fire. A propane tank fuels the burner, and the flow rate of fuel is controlled and measured by the flowmeter. The data acquisition system (DAQ) is the signal generator for the speaker and the data recorder for microphones. See \Cref{fig:manikin_photos}  for details of the manikin.}
\end{figure}

\Cref{fig:schematic1_hrtf_measurement} shows a schematic diagram of the apparatus used in these experiments. The HRTF measurement apparatus was atypical from that used in the literature \cite{gardner1995hrtf}.  The limitations of the fire environment meant that a traditional acoustic manikin (like the KEMAR™) could not be used because of its cost and lack of fire resistance. Instead, a human head-shaped and human-head-sized hollow glass sculpture was used. Two lapel mics (Nady LM-14/U) connected to a wireless transmitter (Nady U41) were installed in a glass manikin head.  This means the absolute HRTF measurement cannot be directly compared with those found in the literature that were made with the KEMAR manikin. However, relative changes in the HRTF can still be measured, which is the purpose of this work. Another difference is that traditional HRTF measurements are typically made with high spatial resolution (source/head elevation and azimuth angles). Because of the transient nature of the fire, measurements in this work are limited to a single source-receiver position for each experiment and thus only two angles are measured relative to the source.

\begin{figure}[]
\centering
\includegraphics[width=0.9\linewidth]{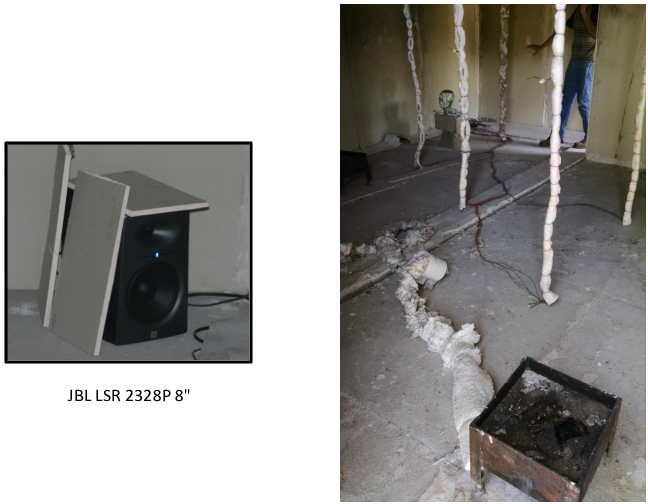}
\caption{\label{fig:hrtf_photograph_inside}
(Color Online)  Photograph of the inside of the burn structure showing the burner, manikin, and speaker used in the HRTF measurement. Several thermocouple trees are visible. The propane line to the sand burner is protected by 2-inch-thick kaowool insulation, and the speaker is protected from the heat using 3/16-inch-thick sheets of gypsum drywall.}
\end{figure}

\begin{figure}[]
\centering
\includegraphics[width=0.9\linewidth]{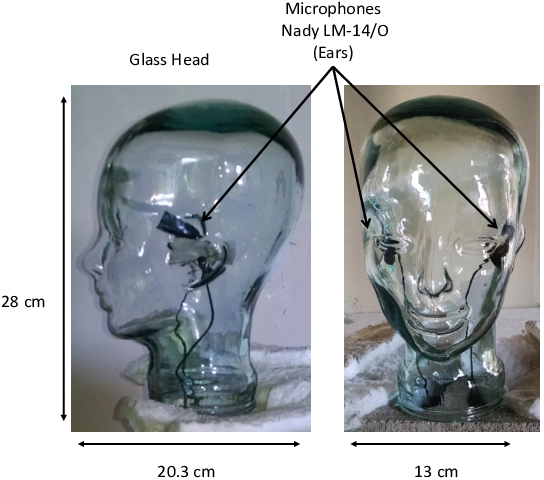}
\caption{\label{fig:manikin_photos}
(Color Online)  This figure shows a manikin head that was used to measure the head-related transfer functions (HRTF) described in \Cref{sec:experiment_2_overview}. The head is made of glass, with lapel microphones as the receivers implanted in the ears. Holes were drilled in the ear locations to install the microphones. The microphones are connected to a wireless transmitter.}
\end{figure}

A JBL LSR2328P (8" two-way Bi-Amplified powered studio monitor) speaker was the acoustic source. A \SI{0.2}{\second}, \SI{100}{\Hz} to \SI{5000}{\Hz} log-frequency modulated signal was generated by an NI-9263 analog output module at \SI{100}{\kHz} sample rate. Simultaneous digitization from the microphones at \SI{10}{\kHz} was performed using a NI-9215 Analog Input Module (BNC). Both the NI-9215 and the NI-9178 were installed in a NI-cDAQ 9178 CompactDAQ chassis. The chassis interfaced with a MATLAB™ based data acquisition script over a USB-2.0 bus. The burner, burn structure, propane source, and measurement setup were all identical to the setup described in \cite{Abbasi2020_Change}. \Cref{fig:hrtf_photograph_inside} shows the experimental setup, and \Cref{fig:manikin_photos} shows the glass manikin head. Six experiments were conducted using this apparatus, with the burner, speaker, and manikin head positions, modeling various PASS rescue scenarios. These experiments are described in \Cref{sec:hrtf_exp_schems}

\subsection{Results}
\label{sec:orgb552dcc}

The evolution of a head-related transfer function in a compartment fire was measured in experiments 3 through 8. These were analyzed in the time and frequency-domain representations. The time-domain HRTF conveys the time delay of arrival between the source and the microphone located at each ear. The time difference of arrival between ear 1 and ear 2 is the interaural time delay (ITD), a key localization cue for low-frequency sound (<1500~Hz) \cite{WightmanFL1992Tdro}.  Frequency-domain HRTF measured the acoustic pressure as a function of frequency at each ear due to unit amplifier input. The ratio of the mean square pressure at ear 1 and ear 2 is the inter-aural level difference (ILD). ILD is a high frequency (>1500~Hz) localization cue. Any changes to the HRTF could affect localization and detection. 

If possible, it would have been instructive to conduct multiple replicates of each experiment.  Given the difficulty of working with real fire, it was only possible to conduct two replicates of experiment 3 and two replicates of experiment 4. \citet[Appendix 1]{abbasi2020Sound} shows the differences between repeated runs of experiments 3 and 4. Differences greater than those observed between repeated measurements with the same conditions can reasonably be assumed to be caused by the different conditions associated with each experiment. The analysis provides evidence that the differences in the HRTF between experiments 3 through 8 are largely real effects caused by differences in the head orientation, fire, and burn structure configurations. They are significantly larger than the differences observed in back-to-back replicates.

\begin{figure}[]
\centering
\includegraphics[angle=0,width=1.0\linewidth, keepaspectratio,,height=0.8\textheight]{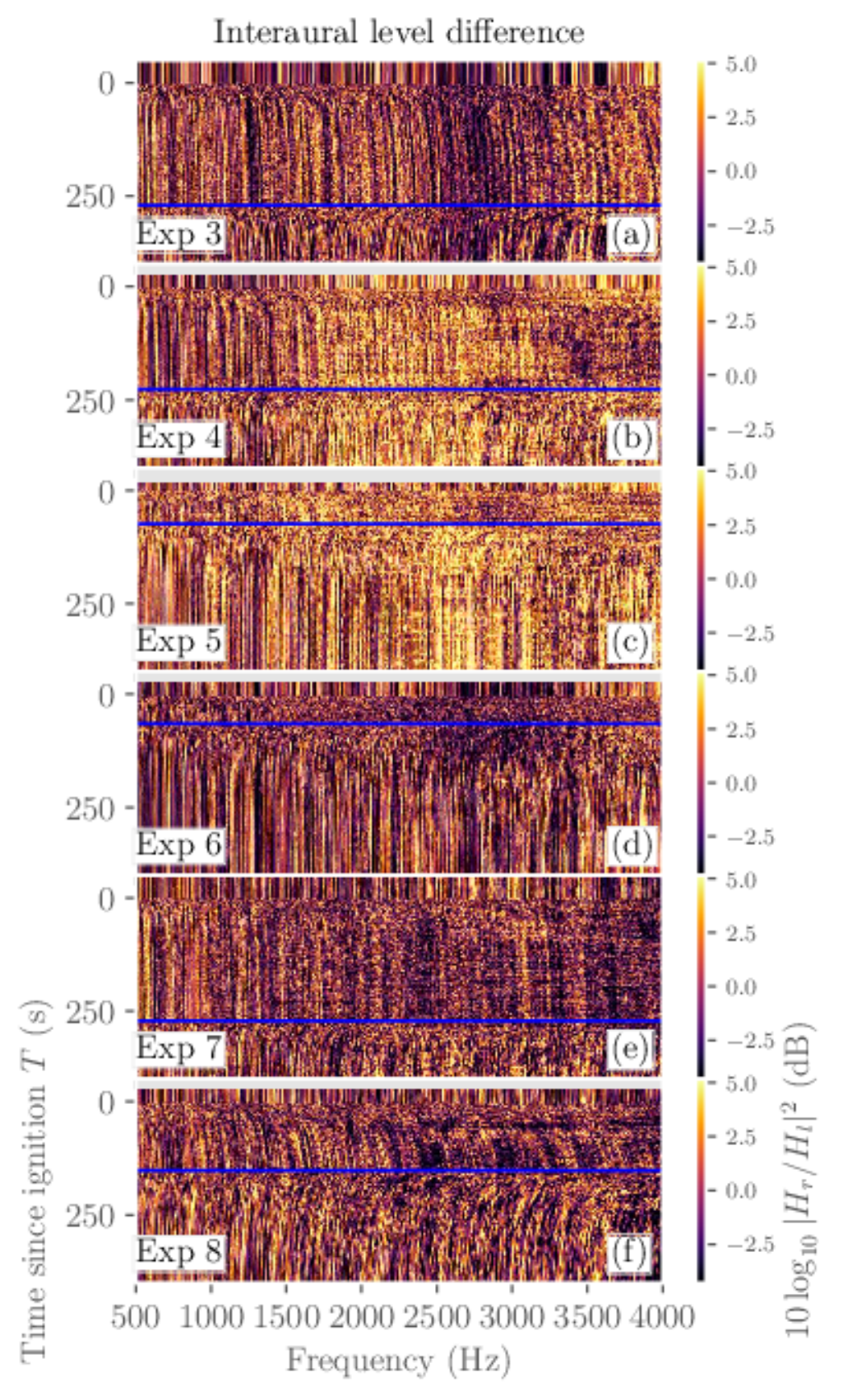}
\caption{\label{fig:Exp_Num_=_23_date_=_2014_08_25_15_26__0_images}
(Color Online)  Measured interaural level difference (ILD) for experiments 3 through 8.}
\end{figure} 

\begin{figure}[]
\centering
\includegraphics[angle=0,width=1.0\linewidth, keepaspectratio ,,height=0.8\textheight]{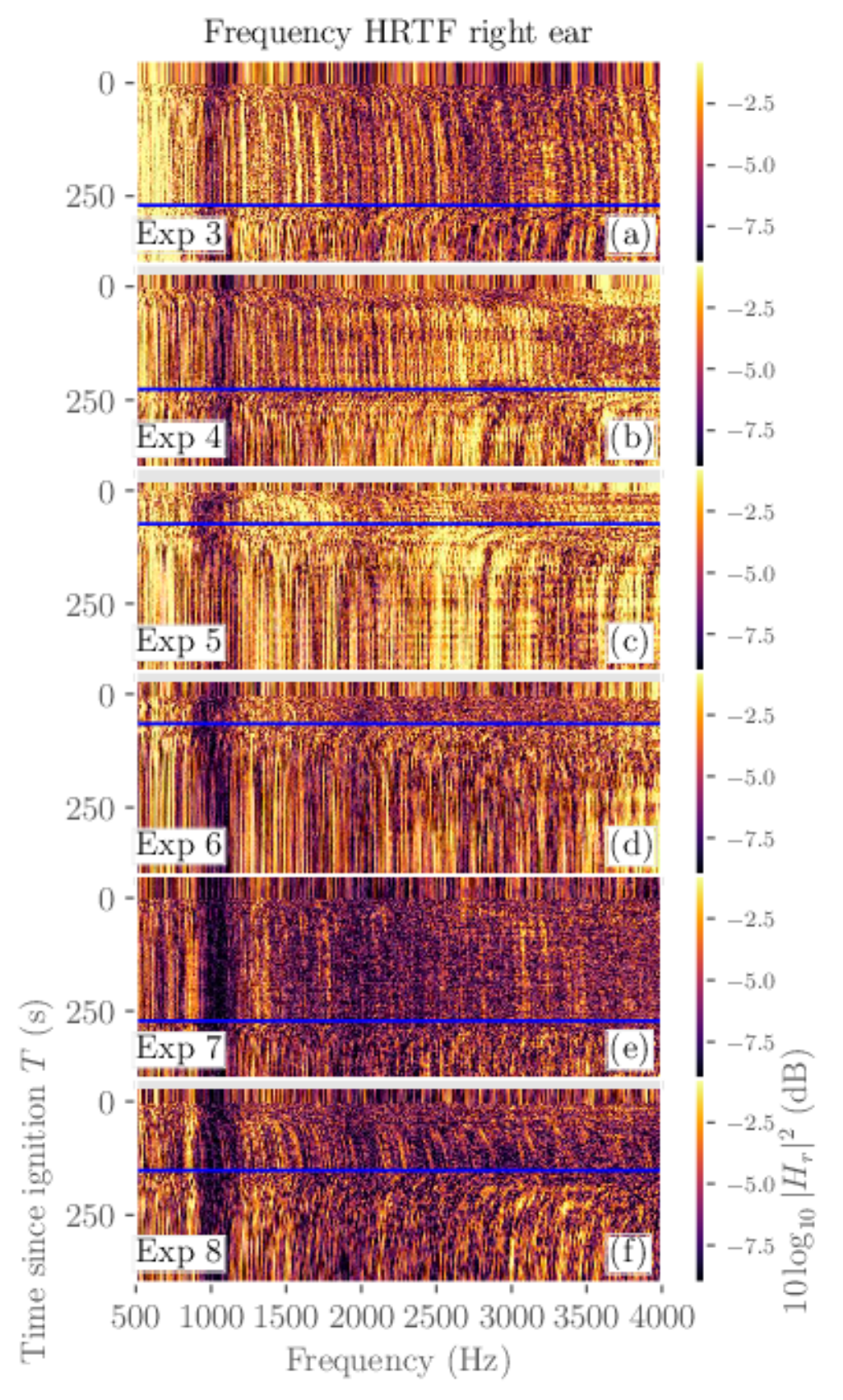}
\caption{\label{fig:Exp_Num_=_23_date_=_2014_08_25_15_26__1_images}
(Color Online)  Measured right `ear' frequency domain head-related transfer function (HRTF) for experiments 3 through 8.}
\end{figure} 

\begin{figure}[]
\centering
\includegraphics[angle=0,width=1.0\linewidth, keepaspectratio ,,height=0.8\textheight]{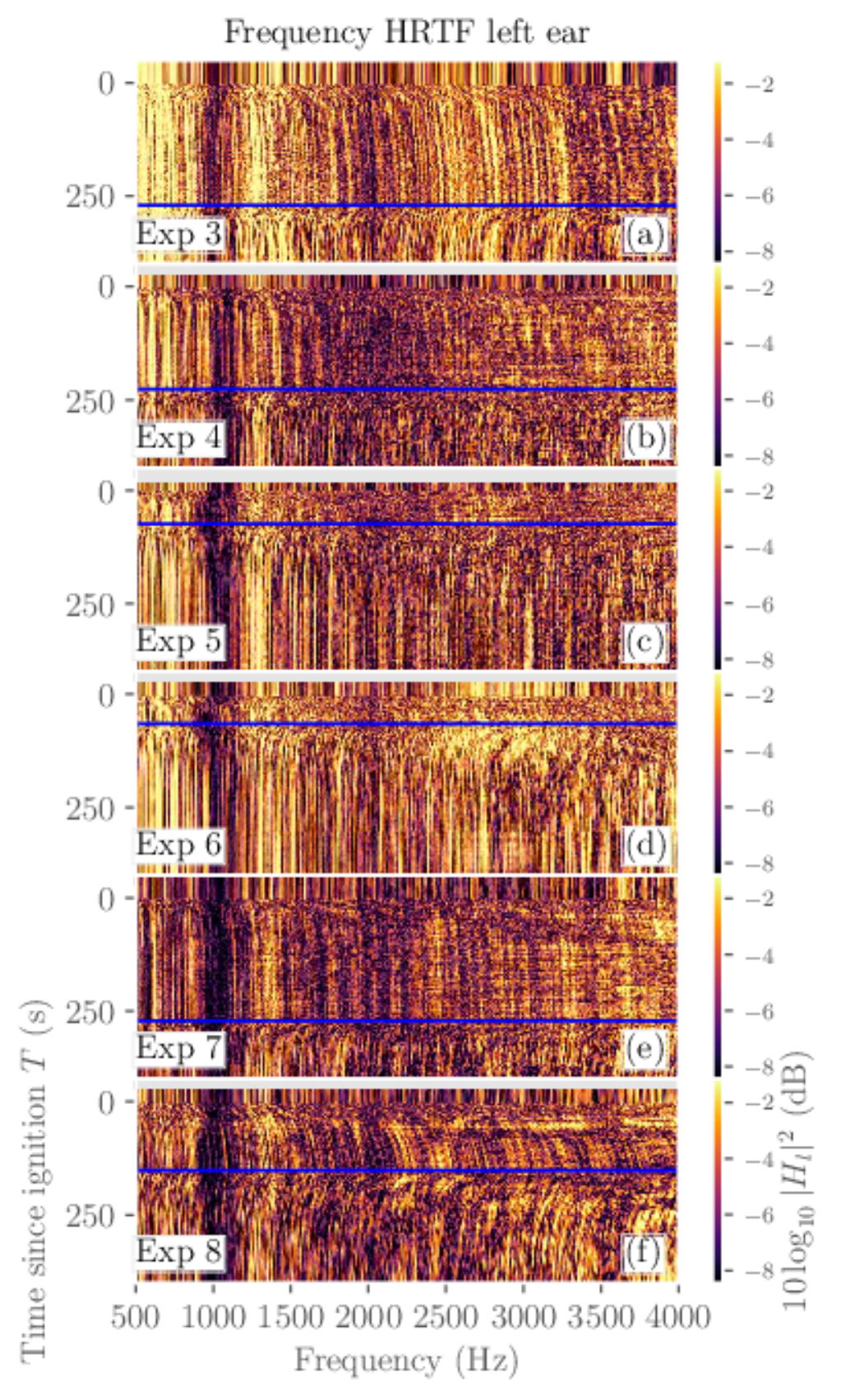}
\caption{\label{fig:Exp_Num_=_23_date_=_2014_08_25_15_26__2_images}
(Color Online)  Measured left `ear' frequency domain head-related transfer function (HRTF) for experiments 3 through 8.}
\end{figure} 

\begin{figure}[]
\centering
\includegraphics[angle=0,width=1.0\linewidth, keepaspectratio ,,height=0.8\textheight]{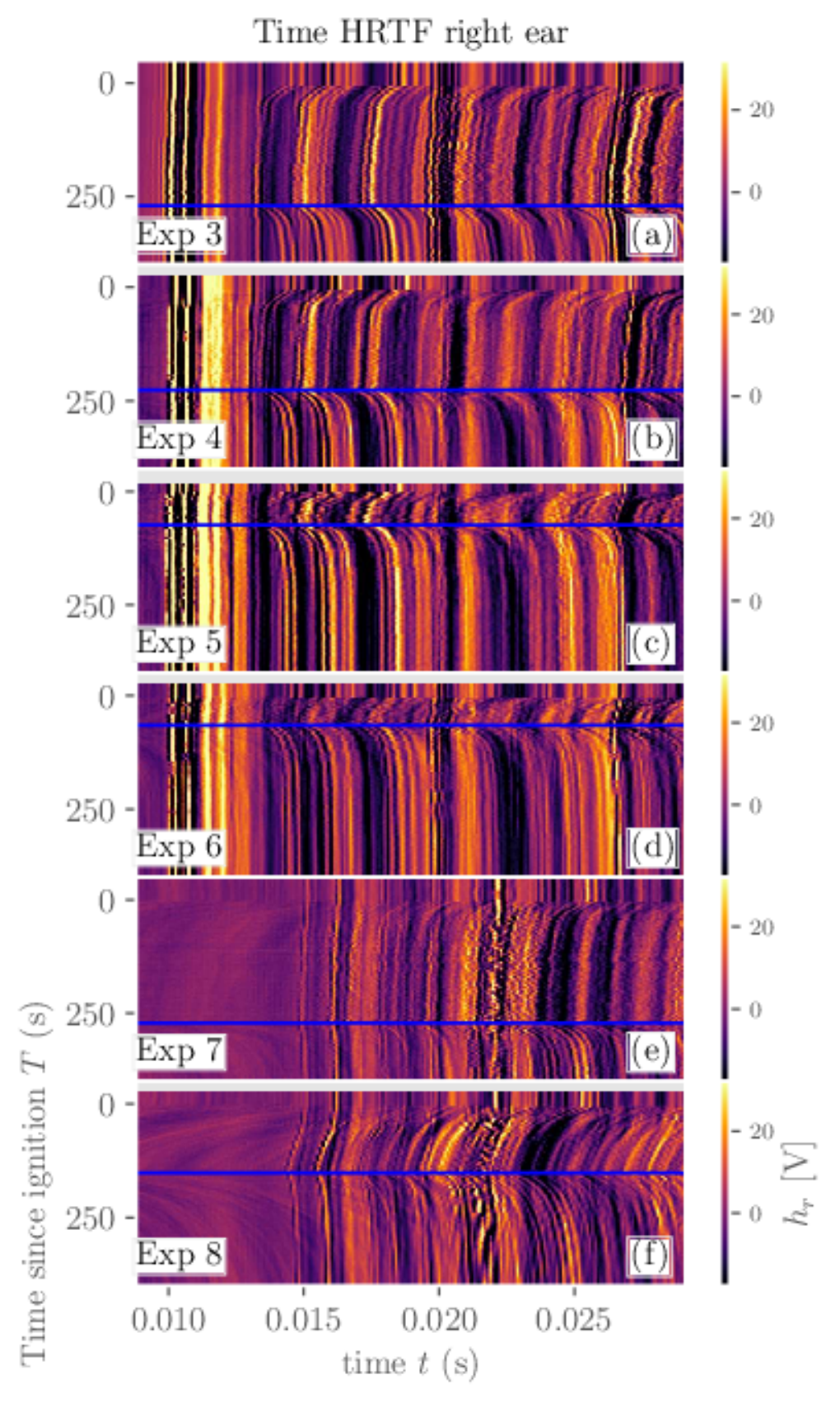}
\caption{\label{fig:Exp_Num_=_23_date_=_2014_08_25_15_26__0_images_time}
(Color Online)  Measured right `ear' time-domain head-related transfer function (HRTF) for experiments 3 through 8.}
\end{figure} 

\begin{figure}[]
\centering
\includegraphics[angle=0,width=1.0\linewidth, keepaspectratio ,,height=0.8\textheight]{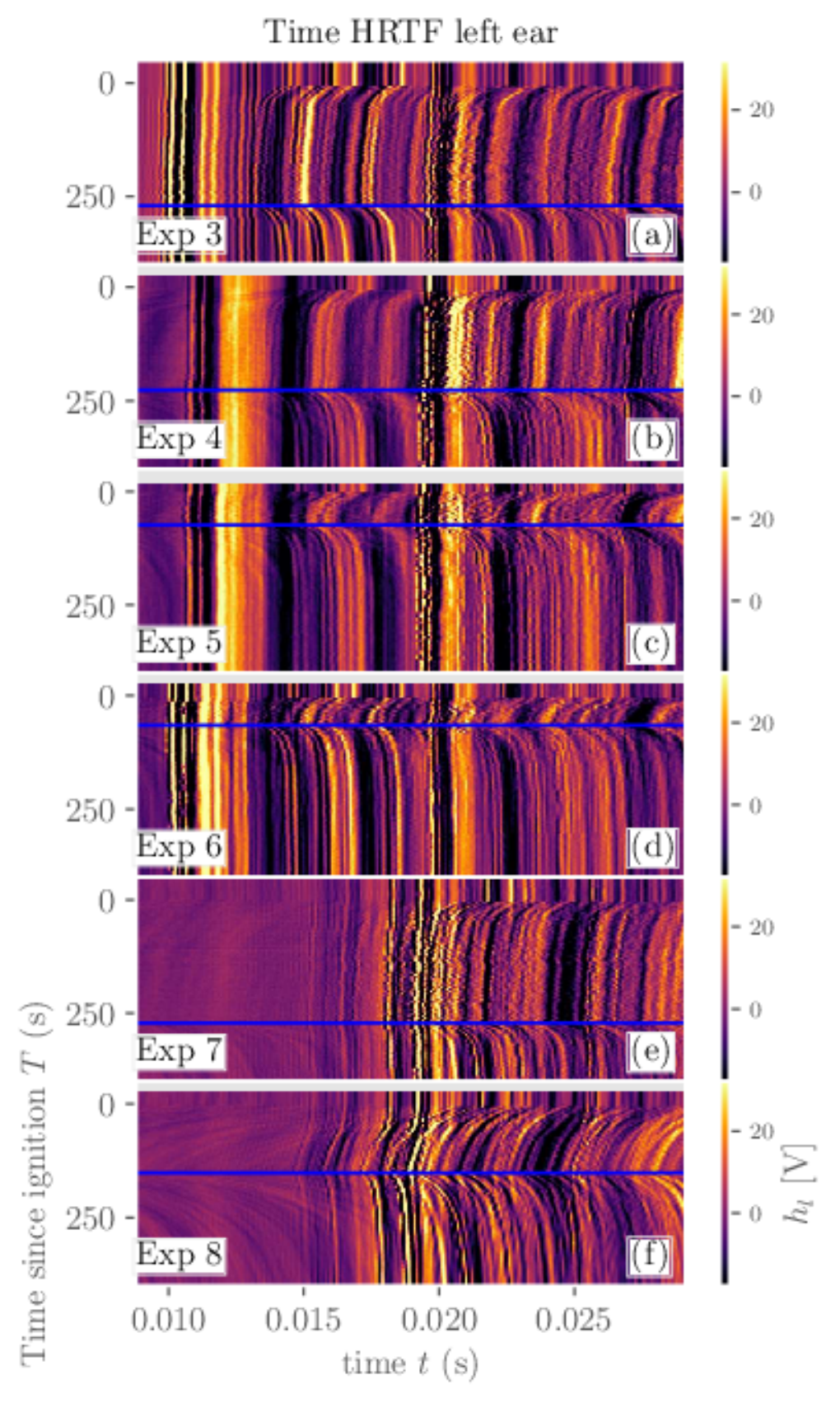}
\caption{\label{fig:Exp_Num_=_23_date_=_2014_08_25_15_26__1_images_time}
(Color Online)   Measured left `ear' time-domain head-related transfer function (HRTF) for experiments 3 through 8.}
\end{figure} 

\begin{figure}[]
\centering
\includegraphics[angle=0,width=1.0\linewidth, keepaspectratio ,,height=0.8\textheight]{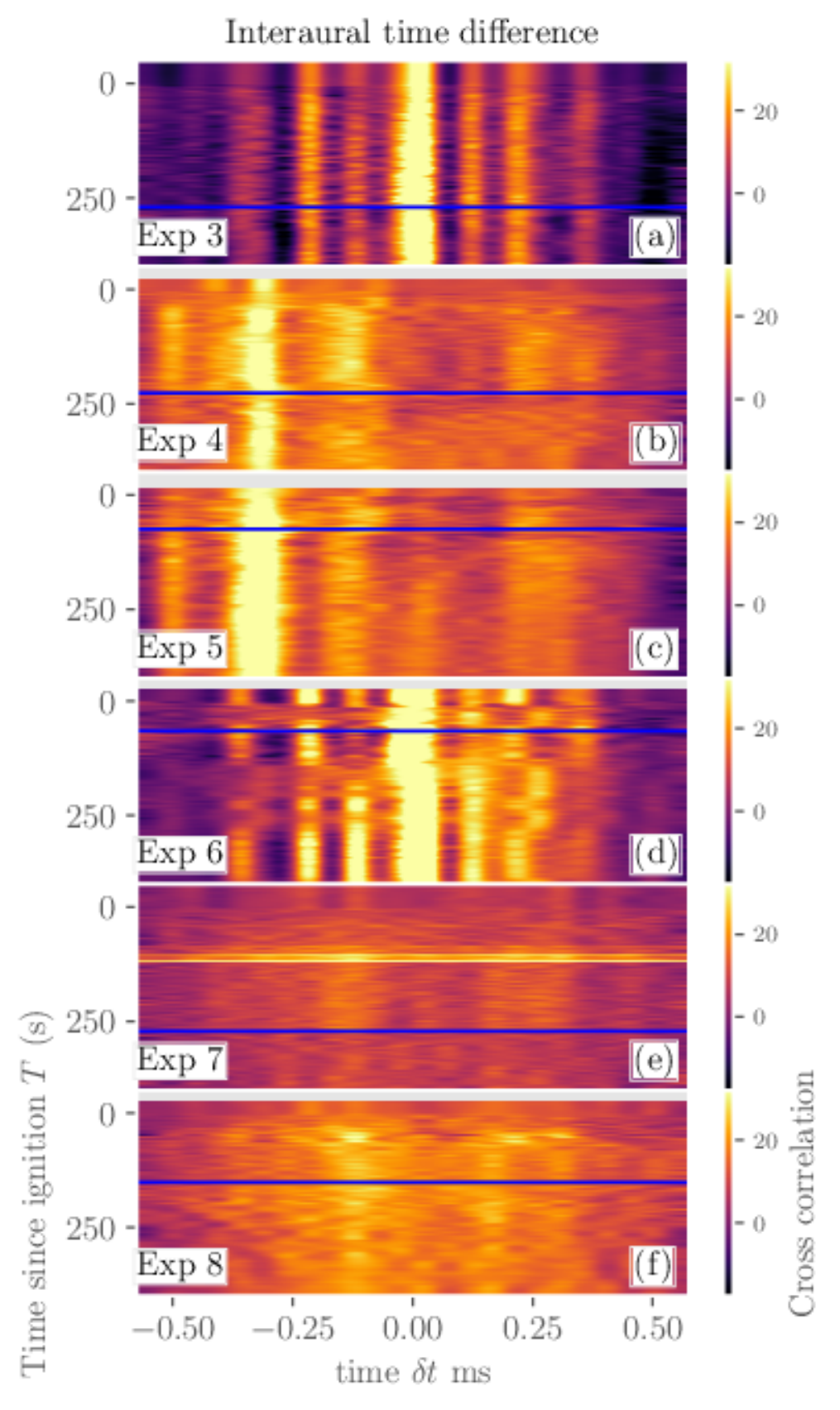}
\caption{\label{fig:Exp_Num_=_23_date_=_2014_08_25_15_26__2_images_time}
(Color Online)   Measured interaural time delay for experiments 3 through 8.}
\end{figure}

Experiments 3 through 8 capture the head-related transfer function (HRTF) between the source and the right and left ears microphones. This is further processed to compute the ILD and ITD. The data is presented in  \Cref{fig:Exp_Num_=_23_date_=_2014_08_25_15_26__0_images}  to \Cref{fig:Exp_Num_=_23_date_=_2014_08_25_15_26__2_images_time}. \Cref{tab:freq_hrtf_figure} and \Cref{tab:time_hrtf_figure} show the key between experiment, measurement type and figure. For example, the interaural time delay for experiment 6 is \Cref{fig:Exp_Num_=_23_date_=_2014_08_25_15_26__2_images}(d). The figures show the evolution of the quantity of interest (HRTF in time, HRTF in frequency, ILD, ITD) as function of time \(T\) after ignition. The blue horizontal line marks the extinction time.

\begin{table}[h]
\caption{Figure key for frequency-domain HRTF and interaural level difference measurements}
\label{tab:freq_hrtf_figure}
\centering
\adjustbox{max width=\linewidth}{
\begin{tabular}{|c|c|c|c|}
\hline
Experiment & Frequency HRTF (right ear) & Frequency HRTF (left ear) & Interaural level difference\\
\hline
3 & \Cref{fig:Exp_Num_=_23_date_=_2014_08_25_15_26__1_images}(a) & \Cref{fig:Exp_Num_=_23_date_=_2014_08_25_15_26__2_images}(a) & \Cref{fig:Exp_Num_=_23_date_=_2014_08_25_15_26__0_images}(a)\\
4 & \Cref{fig:Exp_Num_=_23_date_=_2014_08_25_15_26__1_images}(b) & \Cref{fig:Exp_Num_=_23_date_=_2014_08_25_15_26__2_images}(b) & \Cref{fig:Exp_Num_=_23_date_=_2014_08_25_15_26__0_images}(b)\\
5 & \Cref{fig:Exp_Num_=_23_date_=_2014_08_25_15_26__1_images}(c) & \Cref{fig:Exp_Num_=_23_date_=_2014_08_25_15_26__2_images}(c) & \Cref{fig:Exp_Num_=_23_date_=_2014_08_25_15_26__0_images}(c)\\
6 & \Cref{fig:Exp_Num_=_23_date_=_2014_08_25_15_26__1_images}(d) & \Cref{fig:Exp_Num_=_23_date_=_2014_08_25_15_26__2_images}(d) & \Cref{fig:Exp_Num_=_23_date_=_2014_08_25_15_26__0_images}(d)\\
7 & \Cref{fig:Exp_Num_=_23_date_=_2014_08_25_15_26__1_images}(e) & \Cref{fig:Exp_Num_=_23_date_=_2014_08_25_15_26__2_images}(e) & \Cref{fig:Exp_Num_=_23_date_=_2014_08_25_15_26__0_images}(e)\\
8 & \Cref{fig:Exp_Num_=_23_date_=_2014_08_25_15_26__1_images}(f) & \Cref{fig:Exp_Num_=_23_date_=_2014_08_25_15_26__2_images}(f) & \Cref{fig:Exp_Num_=_23_date_=_2014_08_25_15_26__0_images}(f)\\
\end{tabular}
}
% \hline
\end{table}

\begin{table}[H]
\caption{Figure key for time-domain HRTF and interaural time delay measurements}
\label{tab:time_hrtf_figure}
\centering
\adjustbox{max width=\linewidth}{
\begin{tabular}{|c|c|c|c|}
\hline
Experiment & Time HRTF (right ear) & Time HRTF (left ear) & Interaural time delay\\
\hline
3 & \Cref{fig:Exp_Num_=_23_date_=_2014_08_25_15_26__1_images_time}(a) & \Cref{fig:Exp_Num_=_23_date_=_2014_08_25_15_26__2_images_time}(a) & \Cref{fig:Exp_Num_=_23_date_=_2014_08_25_15_26__0_images_time}(a)\\
4 & \Cref{fig:Exp_Num_=_23_date_=_2014_08_25_15_26__1_images_time}(b) & \Cref{fig:Exp_Num_=_23_date_=_2014_08_25_15_26__2_images_time}(b) & \Cref{fig:Exp_Num_=_23_date_=_2014_08_25_15_26__0_images_time}(b)\\
5 & \Cref{fig:Exp_Num_=_23_date_=_2014_08_25_15_26__1_images_time}(c) & \Cref{fig:Exp_Num_=_23_date_=_2014_08_25_15_26__2_images_time}(c) & \Cref{fig:Exp_Num_=_23_date_=_2014_08_25_15_26__0_images_time}(c)\\
6 & \Cref{fig:Exp_Num_=_23_date_=_2014_08_25_15_26__1_images_time}(d) & \Cref{fig:Exp_Num_=_23_date_=_2014_08_25_15_26__2_images_time}(d) & \Cref{fig:Exp_Num_=_23_date_=_2014_08_25_15_26__0_images_time}(d)\\
7 & \Cref{fig:Exp_Num_=_23_date_=_2014_08_25_15_26__1_images_time}(e) & \Cref{fig:Exp_Num_=_23_date_=_2014_08_25_15_26__2_images_time}(e) & \Cref{fig:Exp_Num_=_23_date_=_2014_08_25_15_26__0_images_time}(e)\\
8 & \Cref{fig:Exp_Num_=_23_date_=_2014_08_25_15_26__1_images_time}(f) & \Cref{fig:Exp_Num_=_23_date_=_2014_08_25_15_26__2_images_time}(f) & \Cref{fig:Exp_Num_=_23_date_=_2014_08_25_15_26__0_images_time}(f)\\
\end{tabular}
}
\end{table}

In experiment 3 the head is facing the speaker and the fire is on the other side of the room. Since both microphones are approximately equidistant from the source, the HRTF for both microphones has a very similar modal structure and change in mode frequency over time. Mode frequency increases after ignition (\(T\)~>~0 s). Low-frequency modes (f~<~\SI{2000}{\Hz}) are temporally stable while the fire is active. High frequency (f~>~2000~Hz) modal structure is much weaker but can be seen. The ILD has the same characteristics as the HRTF. Increase in low-frequency modes and weakening of some higher frequency modes. Since ILD is used for high-frequency localization, this change could affect the perceptual "size" of the room. 

The time-domain HRTF shows a significant change in arrival energy after ignition. The first arrival is minimally affected. Later arrivals show a significant change in arrival time and loss of arrival path. There are random delta in time delays around a mean time for many paths after ignition. This is caused by random time-dependent changes in the sound speed field. The ITD between the `ears' shows small changes for the early arrivals. However, it is important to note that slight changes in the ITD can have a significant impact on sound source localization. 

Modeling the ears as a two-element line array, neglecting all effects of the head, and assuming plane wave propagation, the source direction in the plane of the ear \(\theta\) can be predicted using the ITD and the assumed sound speed \(c\)~=~\SI{343}{\meter\per\second} using \Cref{eq:ear_array_theta}:
\begin{equation}
\label{eq:ear_array_theta}
\theta =\arccos[\textrm{ITD} \frac{c}{d}],
\end{equation} 
where \(d\) is the distance between the ears (or microphones). Changing from \(c\) ITD~=~0.01 m to \(c\) ITD = 0.02~m results in \(\Delta \theta\)~=~\SI{5.8}{\degree}.

In experiment 4 the head, speaker, and fire are in the same position as experiment 3, with the head rotated such that the right ear is facing the speaker. The following observations can be made. The right `ear', which points into the room and towards the fire, appears to lose modal structure above \SI{3300}{\Hz}, while the left ear, which points away from the fire, loses modal structure above \SI{1700}{\Hz}. A similar phenomenon was observed in experiment 1 \cite{Abbasi2020_Change}, where the frequency response of the microphone in the hallway showed less consistent modal structure. We hypothesized this is because the hot air from the fire flows out of the structure through the open door at the end of the hallway. Therefore, the sound speed field in that region is more unsteady because of the flow and mixing of cold and hot air. The frequency regime where the loss of modal consistency occurs is where the interaural level difference is the dominant localization cue. The ILD changes significantly at higher frequencies between experiments 3 and 4. This is the effect of the change in orientation. In an iso-speed environment, the ILD varies smoothly as the head rotates. The fire changes this smoothness, introducing loss of modes and time-varying modal frequency. Since localization is based on the ILD and the relative change in HRTF with aspect angle, we hypothesize the fire could make finding an alarm signal difficult.

In experiments 5 and 6, the speaker and manikin head are in the same position as experiments 3 and 4, but the burner is moved to a location between the head and the speaker.  The head is facing the speaker in experiment 6 and rotated such that the right ear is facing the speaker in experiment 5. The fire was ignited for a shorter period for these two experiments because of the proximity of the fire to the microphones and speaker. The left ear HRTF suffers a significant loss of modal structure at frequency > 1500~Hz while the same happens on the right `ear'  at frequency > 2500~Hz. In both cases, low frequencies are more strongly changed than in experiments 3 and 4. This causes the ILD to be highly chaotic and inconsistent across the entire band. Even at low frequency, there is a loss of modes. Comparing experiments 3 and 4 to 5 and 6 shows that proximity to the fire has a powerful impact on the frequency response and ILD localization cues. The time-domain HRTF also reflects these changes. There is a larger change on the first arrival path in experiments 5 and 6 than in 3 and 4. Some arrivals are entirely lost after ignition, that return immediately after extinction, showing it was the flame that was causing the loss.

In Experiments 7 and 8, the head is moved to the hallway next to the room with the fire. The only difference is that in experiment 8, the burn structure door is closed. The burner is returned to the position in experiment 3. The left ear faces into the room and the right ear faces out. Observe that. In experiment 7, both ``ears'' see a similar acoustic response. ILD is consistent for frequency <~1300~Hz, and loses modal structure above 1600~Hz. This shows that the acoustic environment changes more in the hallway than in the room with the fire. The modal structure in experiment 8 is more consistent over time than in Experiment 7. This is further evidence towards the hypothesis that the flow of hot gas out through the narrow hallway creates a highly chaotic temperature field in the hallway leading to uncorrelated impulse responses. Experiment 8 also shows the largest change in mode frequency, likely because the closed-door resulted in a higher temperature in the burn structure.

\section{Measurement of Spatial and Temporal Temperature Evolution}
\label{sec:org49f8a3f}
\label{sec:exp_temperature_measurement}

\subsection{Measurement setup}
\label{sec:org6331015}
The burn structure was equipped with thermocouples to measure the temperature during the experiment. This section shows the evolution of the temperature field during experiment 3.

\begin{figure}[]
\centering
\includegraphics[width=0.9\linewidth]{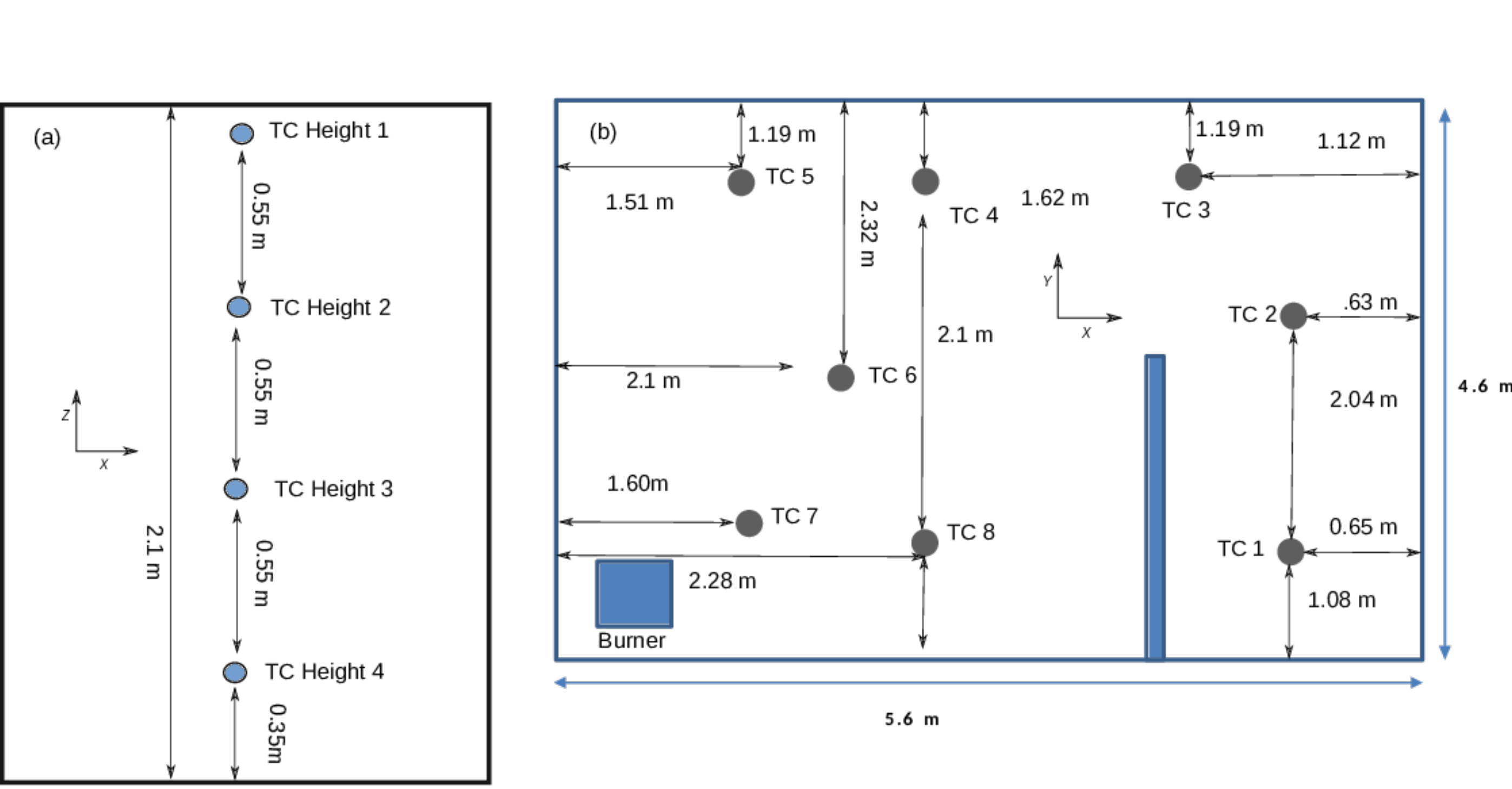}
\caption{\label{fig:thermocouple_tree_schematic}
(Color Online)  Locations of the thermocouples used for the temperature measurement described in \Cref{sec:exp_temperature_measurement}. A side view (a) shows the height of the thermocouples (same in each tree).  A plan view is shown in (b).  See \Cref{fig:hrtf_photograph_inside} for a photograph of several of the trees installed in the burn structure. See \Cref{fig:exp3_schem} for the location of the other equipment in experiment 3.}
\end{figure}

The thermocouples were distributed throughout the burn structure. They were grouped into vertical ``trees,''  hanging from the ceiling and anchored to the floor, and the core of the tree was insulated using 2-inch-thick Kaowool blanket S \cite{foundrykaowool} (ceramic refractory fiber blanket). \Cref{fig:thermocouple_tree_schematic} shows the locations of the thermocouple trees and the height of the thermocouples on the trees. Each tree is labeled TC1 through TC8.

\subsection{Results}
\label{sec:org95a8f53}
\begin{figure}[]
\centering
\includegraphics[width=1.0\linewidth, keepaspectratio,,height=0.8\textheight,]{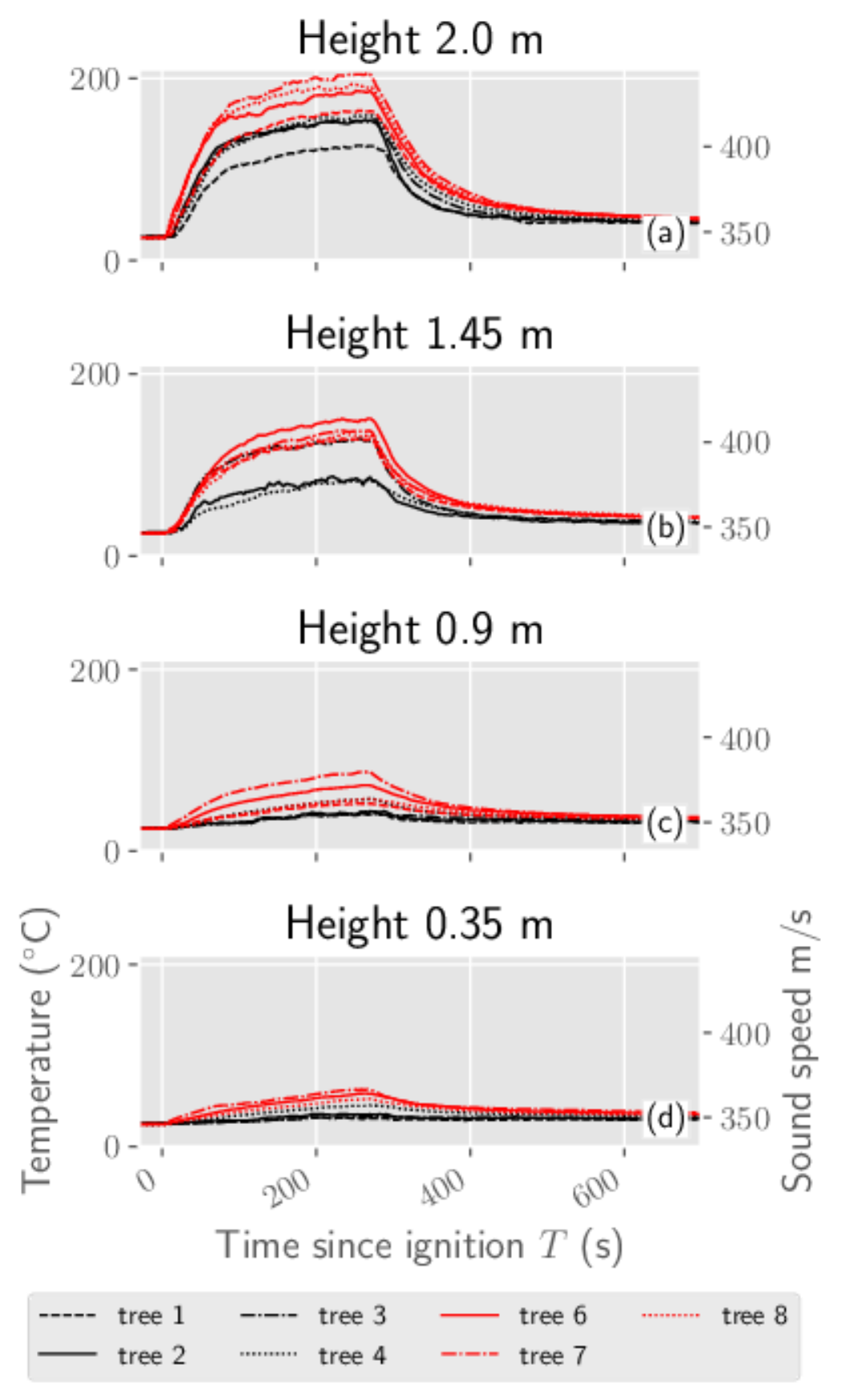}
\caption{\label{fig:tc_temp_plot_1_8}
(Color Online)  Temperature profiles over time organized breproy sensor height for  experiment 3.}
\end{figure}

\Cref{fig:tc_temp_plot_1_8} shows the temperature profiles as measured at each of the eight tree locations.  The maximum temperature in the compartment is near the fire (TC7), with a hot layer reaching a temperature > \SI{200}{\celsius}. There is also a substantial horizontal temperature gradient throughout the compartment. Observe the rapid rise in temperature after ignition, and drop after extinction present at all sensor locations. 

\section{Effect of Fire on Receive Level of a Single Tone}
\label{sec:orgbedabc9}

\begin{figure}[]
\centering
\includegraphics[angle=0,width=1.0\linewidth, keepaspectratio,,height=0.8\textheight]{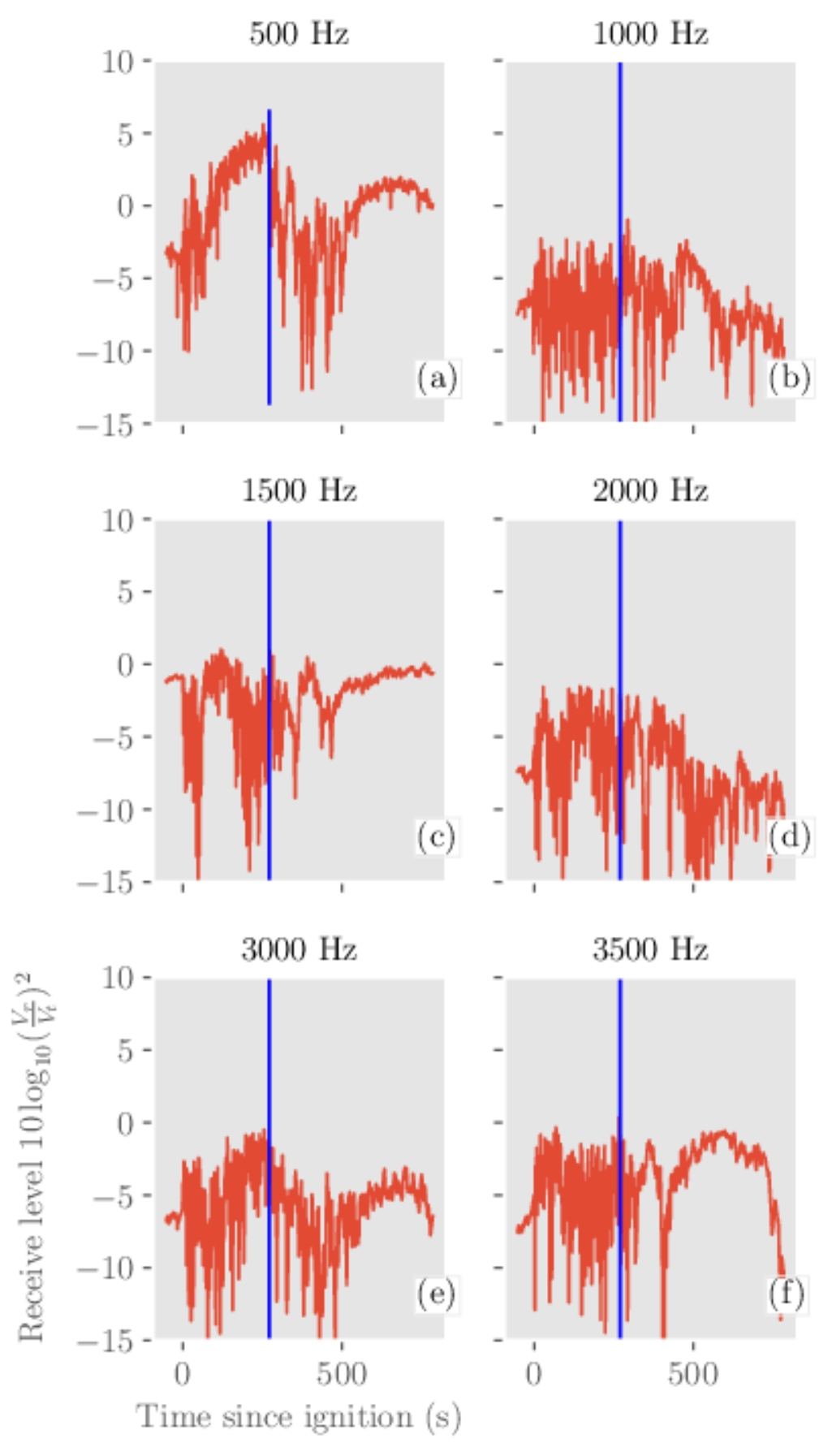}
\caption{\label{fig:Exp_Num_=_23_date_=_2014_08_25_15_26__TXferforfreq}
(Color Online)  Evolution of receive level in the right ``ear'' microphone for select frequencies in experiment 3. Blue vertical line marks the extinction time.}
\end{figure}

\Cref{fig:Exp_Num_=_23_date_=_2014_08_25_15_26__TXferforfreq} shows the received level of 500~Hz, 1000~Hz, 1500~Hz, 2000~Hz, 2500~Hz, 3000~Hz and 3500~Hz tones on the right ear microphone for experiment 3. Ignition is at \(T\)~=~0~s, and the blue vertical line marks the extinction time. The receive level is \(10 \log_{10} (\frac{V_{r}}{V_{t}})^2\), where \(V_r\) is the voltage measure by the receiver and \(V_t\) is the voltage output from the data acquisition system. Before ignition, the received level is steady to within 2~dB. After ignition, the level changes. If a firefighter was relying on a tonal alarm (as in 2007 PASS) the level of the tone would become time-varying due to the fire, potentially confusing localization efforts.

\section{Effect of the Fire on Alarm Signals Heard by a Firefighter}
\label{sec:orgea207bd}
\label{sec:pass_aura}

\begin{figure}[]
\centering
\includegraphics[width=1.0\linewidth, keepaspectratio,,height=0.8\textheight,]{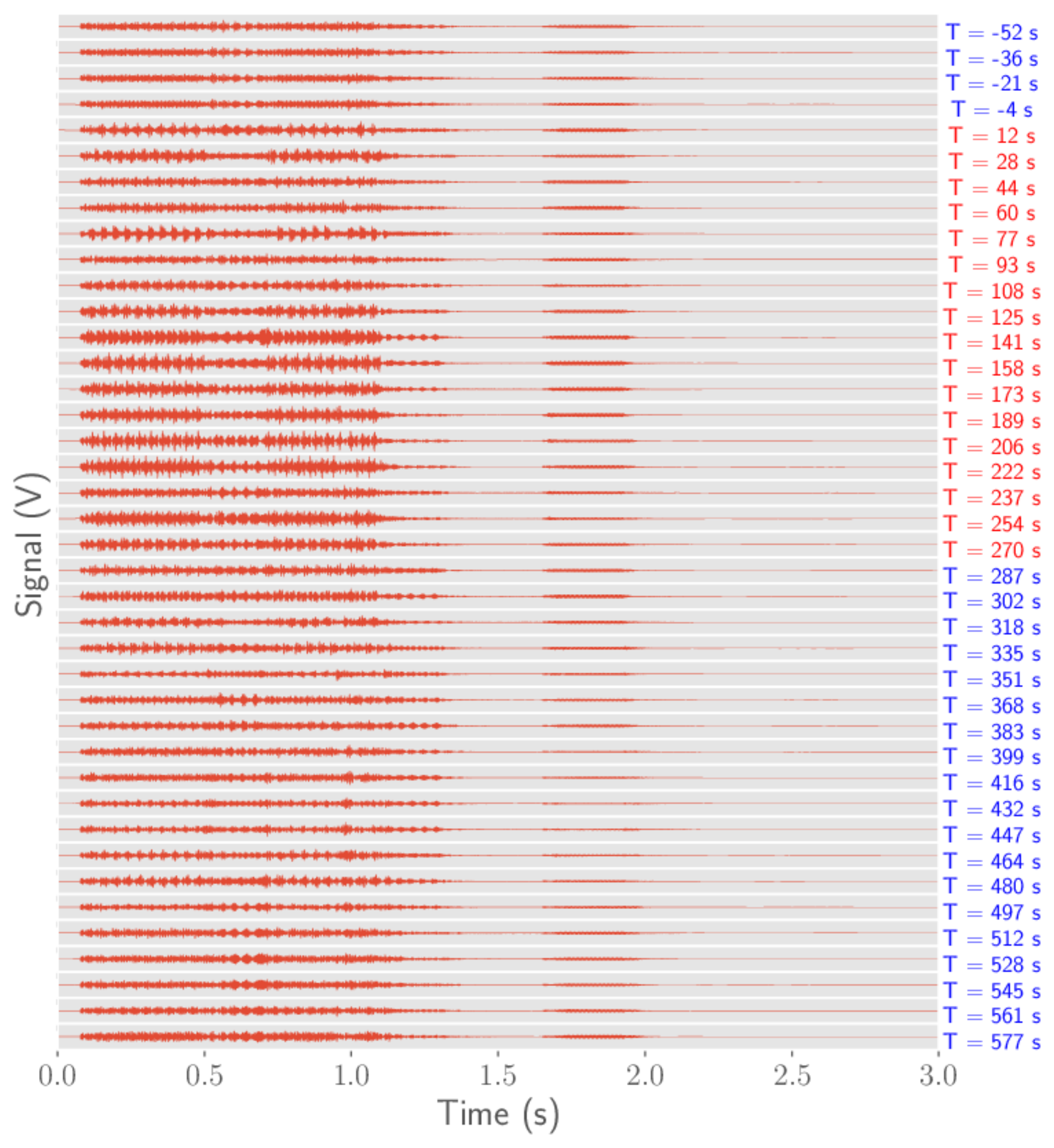}
\caption{\label{fig:2007_pass}
(Color Online) Auralization of the NFPA-1982(2007) PASS alarm signal in the compartment fire for experiment 3 over time. Ignition was at \(T\) = 0. The labels colored red indicate the time when the fire was ignited. The Y-axis scaling of all signals is identical and the signal level can be compared.}
\end{figure}

\begin{figure}[]
\centering
\includegraphics[width=1.0\linewidth, keepaspectratio,,height=0.8\textheight,]{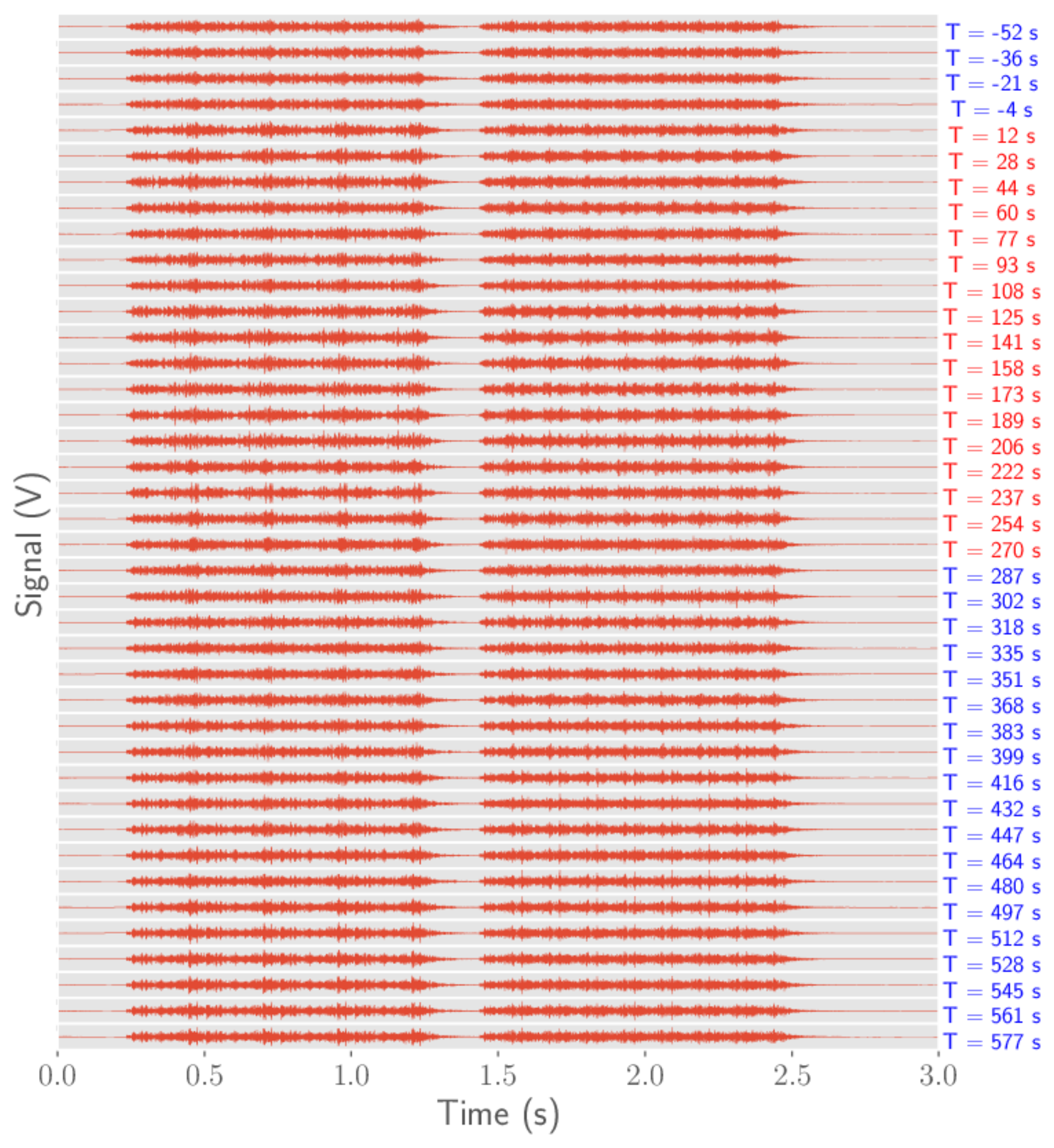}
\caption{\label{fig:2013_pass}
(Color Online) Auralization of the NFPA-1982(2013) PASS alarm signal in the compartment fire for experiment 3 over time. Ignition was at \(T\) = 0. The labels colored red indicate the time when the fire was ignited. The Y-axis scaling of all signals is identical and the signal level can be compared.}
\end{figure}

\Cref{fig:2007_pass,fig:2013_pass} show the auralizations of the 2007 and 2013 PASS alarm signals in the compartment fire for experiment 3 over time. These were computed using the convolution of the measured impulse response with the alarm signal. The way to interpret these auralizations is that they are the sound that would be recorded on a microphone (representing a firefighter's ear) if the PASS alarm was played through a speaker in a compartment fire. Note the change in signal characteristics over time as the fire evolves in the compartment. One of the key takeaways for this work is encapsulated in these figures. The fire has a noticeable impact on the sound heard, significantly altering the waveform. 

\section{Conclusion}
\label{sec:org61b9d73}
This work shows experimental measurements of head-related transfer functions in a room with a fire present.

The head-related transfer function was measured using several different head/source positons. Interaural level difference (ILD) and interaural time delay (ITD) were computed and analyzed. Since ILD and ITD are hearing localization cues, the fire changing them could confuse a listener as to where the sound is coming from. 

Two PASS alarm signals (a 2007 standard and a 2013 standard alarm) were auralized using the measured impulse responses. There was a significant change to the waveform after ignition. Visually it appears that the 2013 standard is more robust to these changes but that cannot be confirmed without human testing. 

This work showed the changes to the acoustics in a room caused by the fire, in the frequency band of interest to the PASS alarm. The various changes are documented, and possible effects on listener performance (detection and localization) were considered. Future work using human subjects is needed to make conclusions on the effect of the fire on the PASS alarm performance.

\section{Acknowledgement}
\label{sec:orga44f2d1}
The experimental measurements were funded by the U.S. Department of Homeland Security Assistance to Firefighters Grants Program. Analysis was self-funded by Dr. Abbasi. The authors thank Mudeer Habeeb, Kyle Ford, and Joelle Suits for their assistance with the experiments.

\section{References}
\label{sec:org70b5081}

\bibliography{Chapter_1_master}

\end{document}